\documentclass[twocolumn,showpacs,prb]{revtex4}
\usepackage{graphicx}
\usepackage[dvips]{color}
\usepackage{amsmath}
\usepackage{multirow}
\newcommand{\be}{\begin{equation}}
\newcommand{\ee}{\end{equation}}
\newcommand{\bea}{\begin{eqnarray}}
\newcommand{\eea}{\end{eqnarray}}
\newcommand{\C}{\mathcal{C}}
\newcommand{\N}{\mathcal{N}}

\begin{document}

\title{Orbital entanglement production in Andreev billiards with time-reversal symmetry}

\author{Sergio Rodr\'{i}guez-P\'{e}rez and Marcel Novaes}
\affiliation{Departamento de F\'{\i}sica, Universidade Federal de S\~ao Carlos, S\~ao
Carlos, SP, 13565-905, Brazil}

\begin{abstract}
We study orbital entanglement production in a chaotic cavity connected to four
single-channel normal-metal leads and one superconducting lead, assuming the presence of
time-reversal symmetry (TRS). The scattered state of two incident electrons is written as
the superposition of several two-outgoing quasi-particle states, four of which are
orbitally entangled in a left-right bipartition. We calculate numerically the mean value
of the squared norm of each scattered state's component, as functions of the number of channels in the superconducting lead. Its behavior is explained as resulting from the proximity effect. We also study statistically the amount of entanglement carried by each pair of outgoing quasi-particles. When the influence of the superconductor is more intense, the average entanglement is found to be considerably larger than that obtained using normal cavities. 
\end{abstract}


\maketitle

\section{Introduction}\label{sec.introduction}

Quantum correlations are in the very heart of quantum physics\cite{EPR} and of
applications like quantum computing and quantum cryptography.\cite{Nielsen} The potential of quantum-transport devices to produce and manipulate entanglement has been explored during the last years.\cite{Beenakker.school,nazarov.livro} Several of those devices produce entanglement from scattering processes taking place in its components.

Chaotic cavities have been proposed as orbital entanglers.\cite{beenakker.emaranhador}
The cavity is connected to four one-channel normal-metal leads, two at the left and two
at the right. An electron leaving the cavity at either the left or the right represents a
qubit, and entanglement between two qubits can be studied considering a left-right
bipartition. The mean value and the variance of the concurrence, a quantifier of
entanglement, were initially calculated in the context of random matrix
theory,\cite{beenakker.emaranhador} followed by more complete statistical
analysis.\cite{Frustalgia,gopar} Non-ideal contacts have also been
considered.\cite{chico,vivo} Constraints on entanglement production imposed by the
geometry of the device were explored,\cite{Sergio-Marcel-normal} and it was found that
more entangled states are less likely to be produced in general.

Scattering processes in normal-superconducting (NS) hybrid systems are another source of
entanglement.\cite{Samuelsson(2003),chinese.NS} Entangled pairs of
quasi-particles can be generated after Andreev reflections which
take place at the normal-superconductor interface. In the present work, we propose the
use of Andreev billiards as orbital entanglers. We study the generation probability of
each pair of quasi-particles, and how this quantity and the amount of entanglement are
affected by the influence of the superconductor. We find a notable increase in the
production of
entanglement induced by the proximity effect,\cite{mcmillan(1968)} when compared to normal cavities. On the other hand, we observe in general a compensation between the amount of entanglement and the probability of producing it. 

The paper is organized as follows. In Sec. \ref{sec.device} we explain how the device is
designed, and what is the structure of its scattering matrix. We also write the scattered
state as a function of the transmission properties of the system. The mean value of the
squared norms of states describing two-outgoing quasi-particles, even those
non-entangled, is analyzed in Sec. \ref{sec.norms}. Statistics of concurrence is studied in Sec.
\ref{sec.concurrence}. We present its full distribution, as well as its mean value and
its variance, for different values of the number of open channels in the superconducting
lead. We summarize and conclude in Sec. \ref{sec.conclusions}. Two additional sections
are included as appendices. The equations used to perform the numerical simulations are
presented in Appendix \ref{sec.simulations}, while some analytical considerations about
the squared norms of the entangled states appear in Appendix \ref{sec.diagrammatics}.

\section{Design of the device}\label{sec.device}

\begin{figure}
\begin{center}
\includegraphics[scale=1,clip]{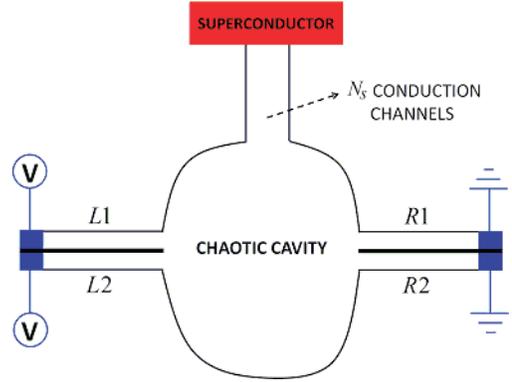}
\end{center}
\caption{(Color online) A chaotic cavity is attached to four normal leads, denoted by
$L_1$, $L_2$, $R_1$ and $R_2$, and one floating superconductor lead. Each normal lead has
one conduction channel while there are $N_s$ open channels in the superconductor lead. A
potential $V$ is applied on the left leads and the right leads are grounded.}
\label{fig.device}
\end{figure}

The setup is represented in Fig. \ref{fig.device}. A central chaotic cavity is connected
to five leads. Four leads are normal and have only one open conduction channel; they are
represented at the left and right sides of the cavity and denoted by $L_1$, $L_2$, $R_1$,
and $R_2$. A superconducting lead with $N_s$ channels is attached to the top of the
cavity. A potential $V$ is applied on the left leads and the right leads are grounded.
The superconducting lead is floating. Thus an electrical current crosses the device by
the normal leads. Specifically, pairs of electrons with energies between the Fermi energy
$E_F$ and $E_F+eV$ enter the cavity from the left leads. It is assumed that the energy of
the incident electrons is much smaller than the gap $\Delta$ in the superconductor,
i.e. $eV\ll\Delta$. This implies that charge transfer in the NS interface takes place only
via Andreev reflection.\cite{andreev} Since the device has only one superconducting lead,
the phase of the superconductor is irrelevant. We also neglect temperature fluctuations.

\subsection{Scattering matrix of the system}\label{subsec.scattering.matrix}

The potential $V$ is small enough to neglect the dependence on energy of the scattering
matrix. We emphasize that this matrix does not belong to any universal class of random
matrices. In order to reach universality in the presence of TRS, it would be necessary to
attach another superconducting lead to the cavity, with a phase difference of
$\pi$.\cite{altland(1997),melsen(1997)} Also the number of open channels in the
superconducting lead must be larger than one.\cite{brouwer(2000)} We focus on a different
regime, where proximity effects are important. The scattering matrix will be denoted by
$S_{NS}$ and has the following structure:


\be
    S_{NS}=\left(
    \begin{array}{cc}
        \hat{r} & \hat{t}^{\prime}\\
        \hat{t} & \hat{r}^{\prime}
    \end{array}
    \right).
    \label{eq.Sns.matrix}
\ee The blocks $\hat{r}$ ($\hat{r}^{\prime}$) and  $\hat{t}$ ($\hat{t}^{\prime}$)
contain, respectively, the reflection and the transmission amplitudes for quasi-particles
coming from the left (right). Each block has its own electron-hole structure, which is
indicated with the hat symbol. For example, the matrix $\hat{r}$ is \be
    \hat{r}=\left(
    \begin{array}{cc}
        r^{ee} & r^{eh}\\
        r^{he} & r^{hh}
    \end{array}
    \right).
    \label{eq.r.matrix}
\ee On the other hand, each sub-block $r^{\alpha\beta}$ in the last equation is a $2
\times 2$ matrix, whose elements are the reflection amplitudes for the leads $L_1$ and
$L_2$. They are given by \be
    r^{\alpha\beta}=\left(
    \begin{array}{cc}
        r^{\alpha\beta}_{11} & r^{\alpha\beta}_{12}\\
        r^{\alpha\beta}_{21} & r^{\alpha\beta}_{22}
    \end{array}
    \right).
    \label{eq.alpha_beta.matrix}
\ee The other blocks of $S_{NS}$ have the same structure. The scattering matrix of the
whole system can be expressed as a function of the scattering matrix of the normal
cavity,\cite{rev.beenakker} which belongs to the Circular Orthogonal Ensemble. Equations
showing this dependence are presented in Appendix \ref{sec.simulations}.

\subsection{Scattered state}\label{subsec.scattered.state}

The scattered state of two incident electrons can be expressed as $|
\Psi_{\mathrm{scat}}\rangle = | \Psi_{LL}\rangle + | \Psi_{RR}\rangle +
|\Psi_{LR}\rangle$. State $| \Psi_{LL}\rangle$ ($| \Psi_{RR}\rangle$) characterizes two
quasi-particles being scattered to the left (right), so is not entangled. One
quasi-particle scattered to the left and the other one to the right is represented by the
state $|\Psi_{LR}\rangle$. In order to give explicit expressions for these three states,
we define the projector matrix \be \sigma^\prime=\mathrm{i}\,\sigma_y\otimes
\begin{pmatrix}
1 & 0\\
0 & 0
\end{pmatrix}.\label{eq.projector.matrix}
\ee The structure of this matrix accounts for the fact that the incident state is given
by two electrons coming from the left. It is also convenient to define the operators
$\mathcal{L}^{\alpha}_{j}$ and $\mathcal{R}^{\alpha}_{j}$, which create a quasi-particle
of type $\alpha$ going out at the lead $L_{j}$ and $R_{j}$, respectively, with energy
$\epsilon$ if $\alpha=e$ and $-\epsilon$ if $\alpha=h$ ($E_F<\epsilon<E_F+eV$). We also
define
\begin{align}
\hat{\delta}=\hat{r} \sigma^\prime \hat{r}^{\,T},\label{eq.delta.matrix}
\\
\hat{\eta}=\hat{t} \sigma^\prime \hat{t}^{\,T},\label{eq.eta.matrix}
\\
\hat{\gamma}=\hat{r}\sigma^\prime \hat{t}^{\,T},\label{eq.gamma.matrix}
\end{align}
where the superscript ``$T$'' indicates the transposition operation. Using the above
objects, the three states can be expressed in the following
way:\begin{align} |\Psi_{LL}\rangle= \frac{1}{2}
\sum\limits_{jk;\alpha\beta}
{\delta_{jk}^{\alpha\beta}\;\mathcal{L}^{\alpha}_{j}\;\mathcal{L}^{\beta}_{k}\;|0\rangle},\label{eq.states.LL}
 \\
| \Psi_{RR}\rangle= \frac{1}{2} \sum\limits_{jk;\alpha\beta}
{\eta_{jk}^{\alpha\beta}\;\mathcal{R}^{\alpha}_{j}\;\mathcal{R}^{\beta}_{k}\;|0\rangle},\label{eq.states.RR} \\
| \Psi_{LR}\rangle= \sum\limits_{jk;\alpha\beta}
{\gamma_{jk}^{\alpha\beta}\;\mathcal{L}^{\alpha}_{j}\;\mathcal{R}^{\beta}_{k}\;|0\rangle},\label{eq.states.LR}
\end{align}
where $|0\rangle$ is the state without electronic excitations. Each term in these sums
represents a pair of quasi-particles of types $\alpha$ and $\beta$ getting out at the
leads $j$ and $k$. The equations presented in this section are general and do not specify
any particular property of the NS system.

\section{Statistics of the squared norms}\label{sec.norms}

\begin{figure}
\begin{center}
\includegraphics[scale=1,clip]{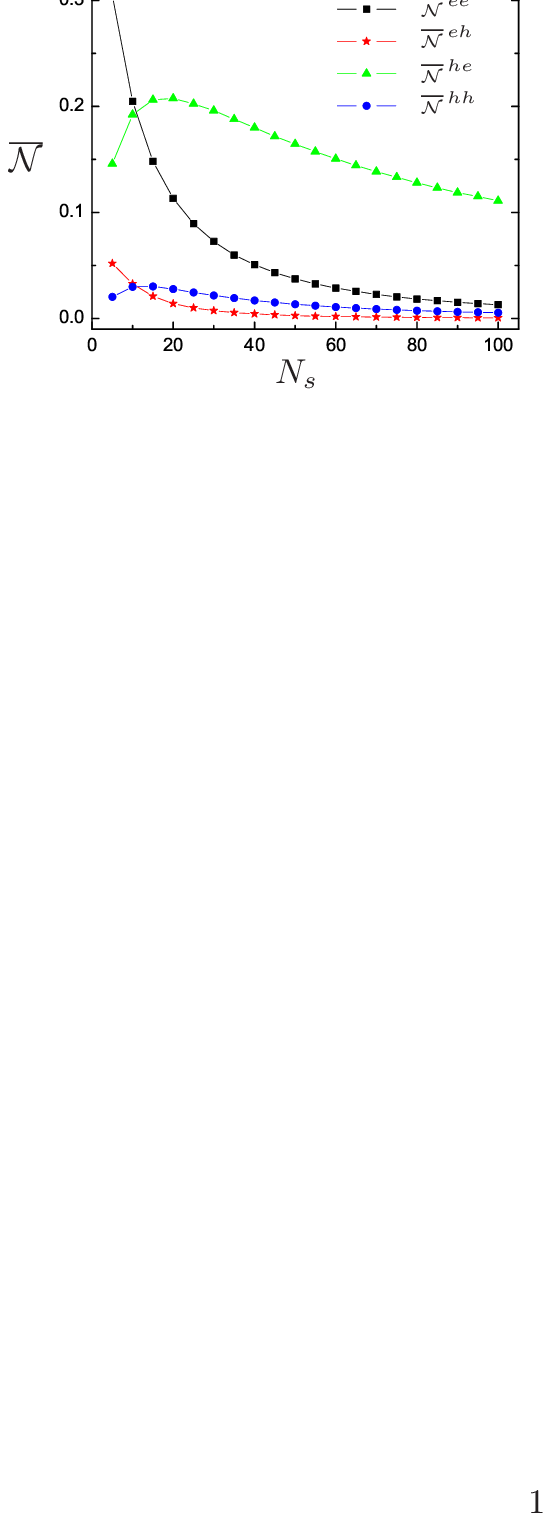}
\includegraphics[scale=1,clip]{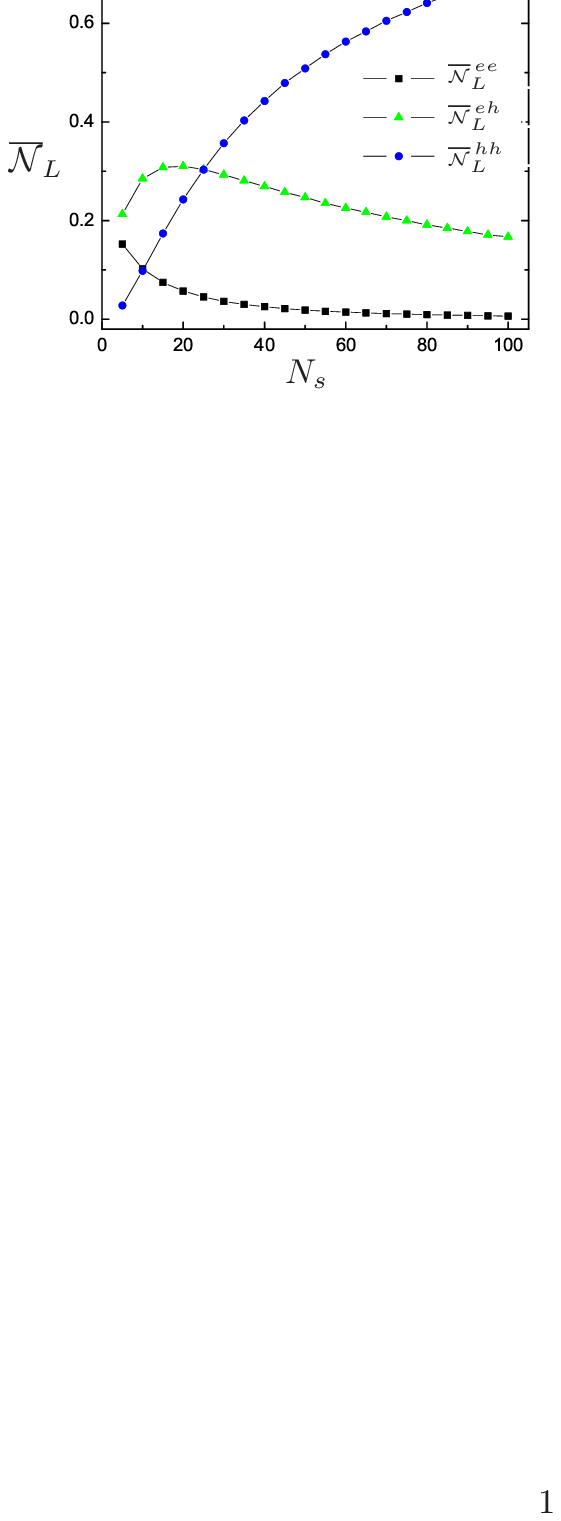}
\includegraphics[scale=1,clip]{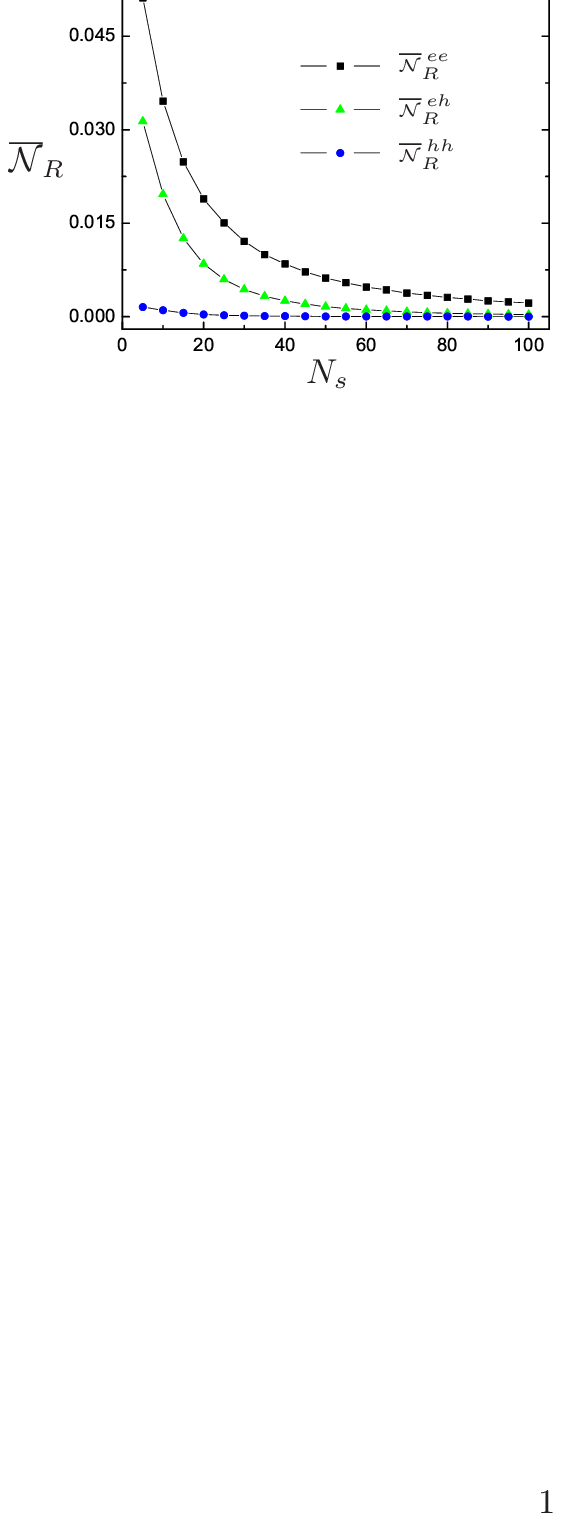}
\end{center}
\caption{(Color online) Mean values of the squared norms $\mathcal{N}^{\alpha\beta}$,
$\mathcal{N}_{L}^{\alpha\beta}$, and $\mathcal{N}_{R}^{\alpha\beta}$ are
represented in panels a), b), and c) respectively, for different numbers of open channels
in the superconducting lead. $10^{5}$ samples were generated for each value of $N_s$.}
\label{fig.norms}
\end{figure}

\begin{figure}
\begin{center}
\includegraphics[scale=1.1,clip]{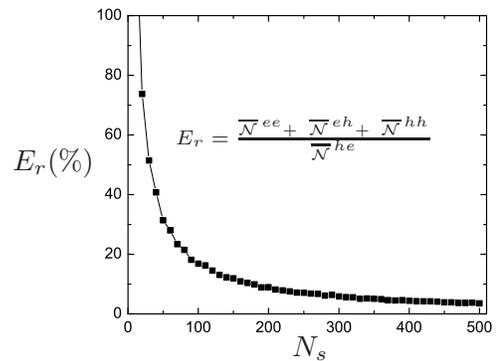}
\end{center}
\caption{Function $E_r$ (in percent) for values of $N_s$ varying from $10$ to $500$. Each
dot was calculated using $5\times10^{3}$ random matrices.} \label{fig.Er}
\end{figure}

In this section we focus on the probability of producing, during a fixed time, a given number of pairs $\alpha$ and $\beta$ being scattered to the left, to the right, or to
different sides. This probability is determined by the squared norm
of the state which represents each process. Let us denote such squared norms as
$\mathcal{N}_{L}^{\alpha\beta}$, $\mathcal{N}_{R}^{\alpha\beta}$, and
$\mathcal{N}^{\alpha\beta}$. We are assuming that $\mathcal{N}^{\alpha\beta}$ refers to
the quasi-particle $\alpha$ scattered to the left and the other one, $\beta$, to the
right. For the other two squared norms there is no distinction between $\alpha,\beta$.
Using Eqs. (\ref{eq.states.LL})-(\ref{eq.states.LR}), they are expressed as functions of the
transmission properties of the system through the following equations:
\begin{align}
\mathcal{N}_{L}^{\alpha\beta}=\mathrm{Tr}(\delta^{\alpha\beta}{\delta^{\alpha\beta}}^\dagger),\label{eq.norms.definitions.LL}\\
\mathcal{N}_{R}^{\alpha\beta}=\mathrm{Tr}(\eta^{\alpha\beta}{\eta^{\alpha\beta}}^\dagger),\label{eq.norms.definitions.RR}\\
\mathcal{N}^{\alpha\beta}=\mathrm{Tr}(\gamma^{\alpha\beta}{\gamma^{\alpha\beta}}^\dagger).\label{eq.norms.definitions.LR}
\end{align}

Let us discuss the underlying mechanism behind the influence of proximity effects on
transport properties. As the number of open channels in the superconducting lead
increases, the probability of incident electrons to undergo an Andreev reflection in the
NS interface and thus to be backscattered as holes to the left before ``feeling'' the
chaos of the cavity gets larger. As $N_s\rightarrow \infty$ this probability becomes
unit, and the device works as if all electrons were Andreev reflected upon arrival at the
cavity.\cite{brouwer(2000)}


A consequence of \emph{direct Andreev backscattering} is an increased number of holes
leaving the cavity at the left, and therefore a decreased transmission of any
quasi-particle to the right. This is verified in Fig. \ref{fig.norms}. As $N_s$ gets bigger, $\overline{\mathcal{N}}^{ee}$, $\overline{\mathcal{N}}^{eh}$,
$\overline{\mathcal{N}}_L^{ee}$, $\overline{\mathcal{N}}_R^{ee}$, and
$\overline{\mathcal{N}}_R^{eh}$ monotonically decrease, because the probability of an
electron to be scattered to the left continually decreases. On the other hand,
$\overline{\mathcal{N}}^{he}$, $\overline{\mathcal{N}}^{hh}$ and
$\overline{\mathcal{N}}_L^{eh}$ have a non-monotonic behavior, initially increasing due
to direct Andreev backscattering. The quantity $\overline{\mathcal{N}}^{he}$ decreases
more slowly than the other squared norms of entangled states because scattering to the
right is more probable for electrons than for holes. Furthermore,
$\overline{\mathcal{N}}_L^{hh}$ ($\overline{\mathcal{N}}_R^{hh}$) monotonically increases
(decreases), also because of direct Andreev backscattering.

The dominance of $\overline{\mathcal{N}}^{he}$ over the other entangled components can be
quantified by \be E_r=\frac{\overline{\mathcal{N}}^{ee}+\;\;
\overline{\mathcal{N}}^{eh}+\;\;\overline{\mathcal{N}}^{hh}}{
\overline{\mathcal{N}}^{he}}.\label{eq.Er} \ee The behavior of $E_r$ is shown in Fig.
\ref{fig.Er}, where $N_s$ runs from $10$ to $500$. Notice that for $N_s=500$, the sum of
the other components represents less than $5\%$ of $\overline{\mathcal{N}}^{he}$.
Although all four kinds of events are rare for large values of $N_s$, a setup for
detection of entanglement would characterize correlations between one left-outgoing hole
and one right-outgoing electron with a small relative error.

Direct Andreev backscattering does not occur if TRS is broken, because
a random phase is accumulated during the process and so it is killed after averaging. An
implication of this is that in the presence of a magnetic field and for $N_s\gg1$, the
statistical distributions of $\mathcal{N}^{\alpha\beta}$ are very similar for any
$\alpha$ and $\beta$. We corroborated this observation through numerical simulations (not
shown).

\section{Statistics of concurrence}\label{sec.concurrence}

\begin{figure}
\begin{center}
\includegraphics[scale=1,clip]{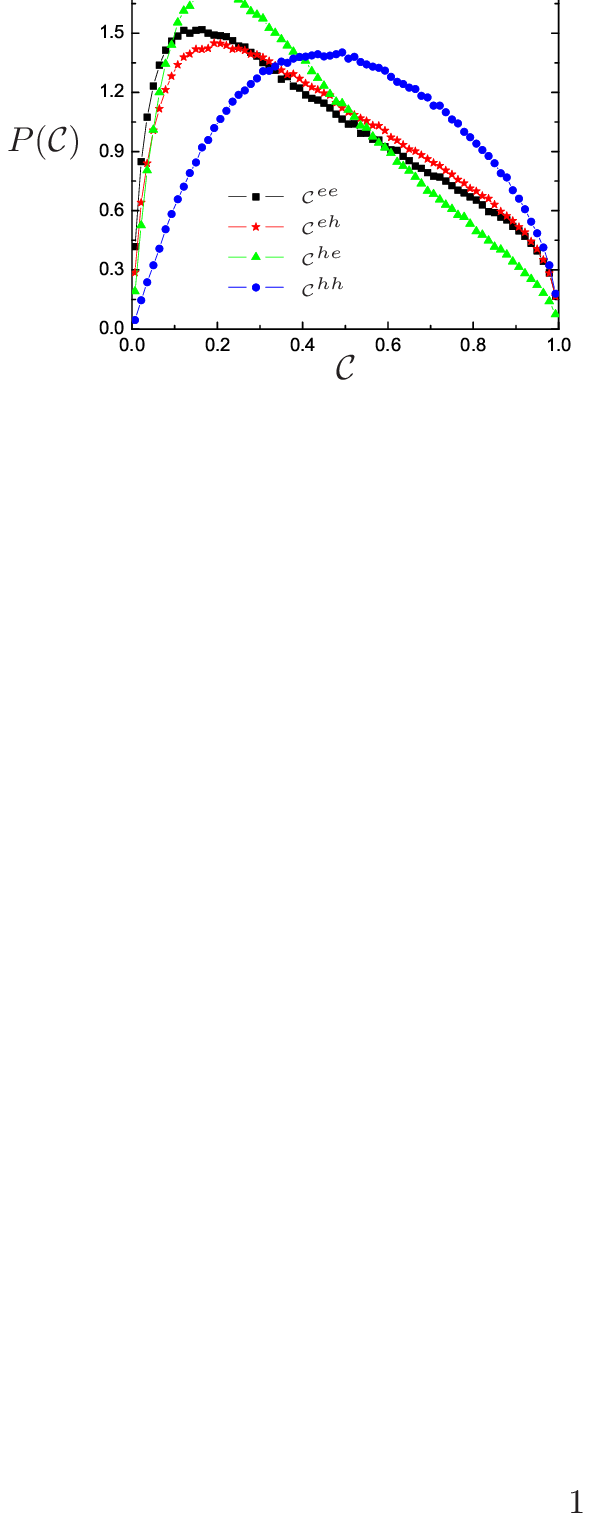}
\includegraphics[scale=1,clip]{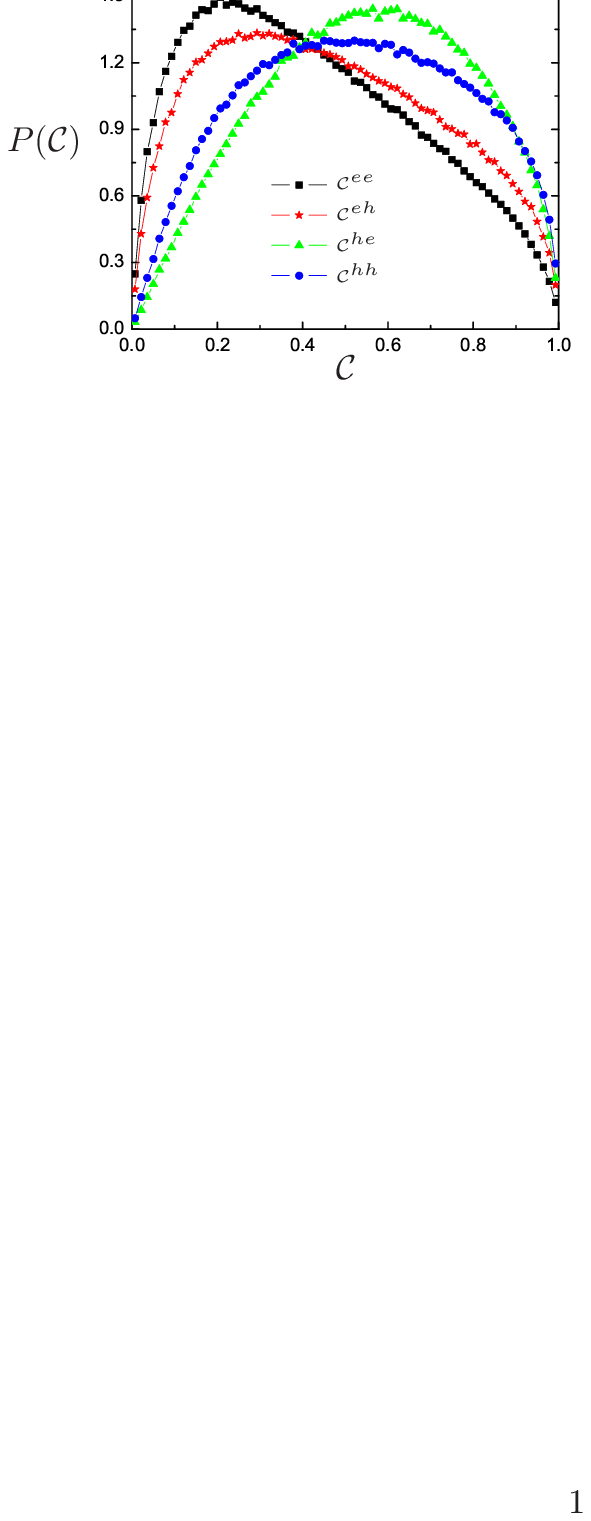}
\includegraphics[scale=1,clip]{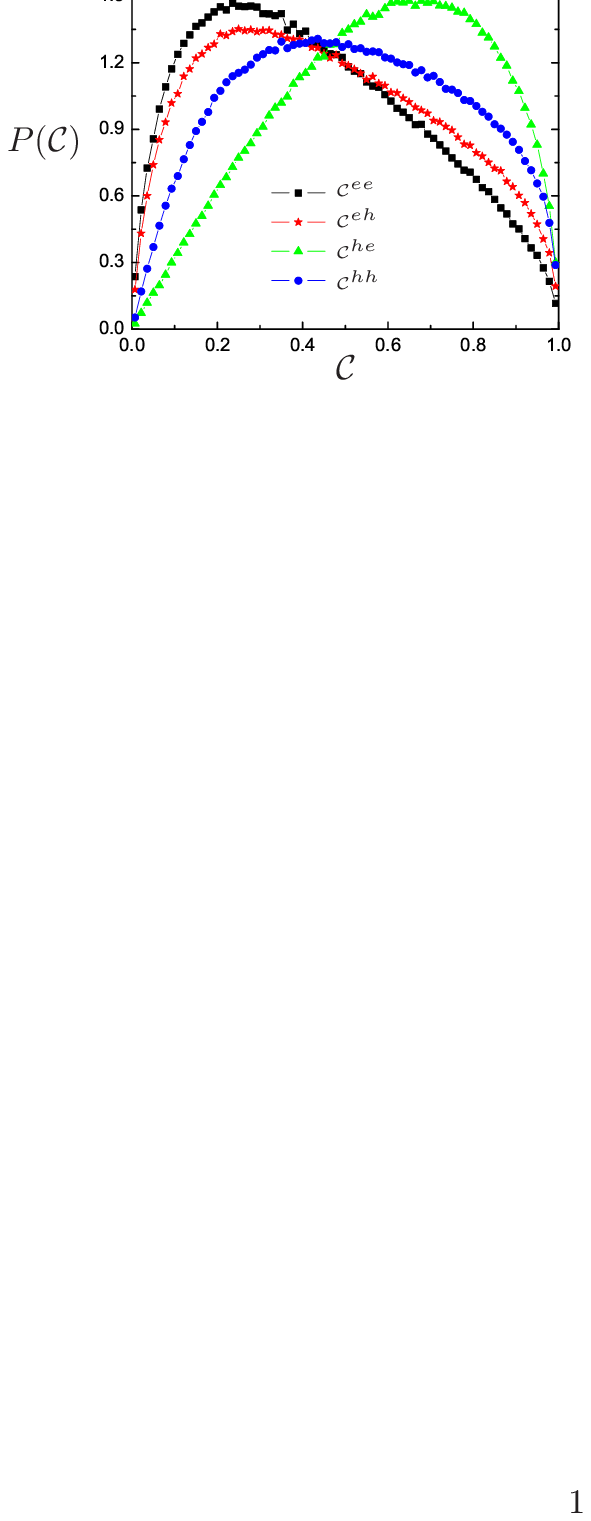}
\end{center}
\caption{(Color online) Distributions of concurrence are represented for $N_s$ equal to
a) $5$, b) $25$, and c) 100. Each distribution was made using $10^6$ random matrices.}
\label{fig.dist.concurrence}
\end{figure}

\begin{figure}
\begin{center}
\includegraphics[scale=1,clip]{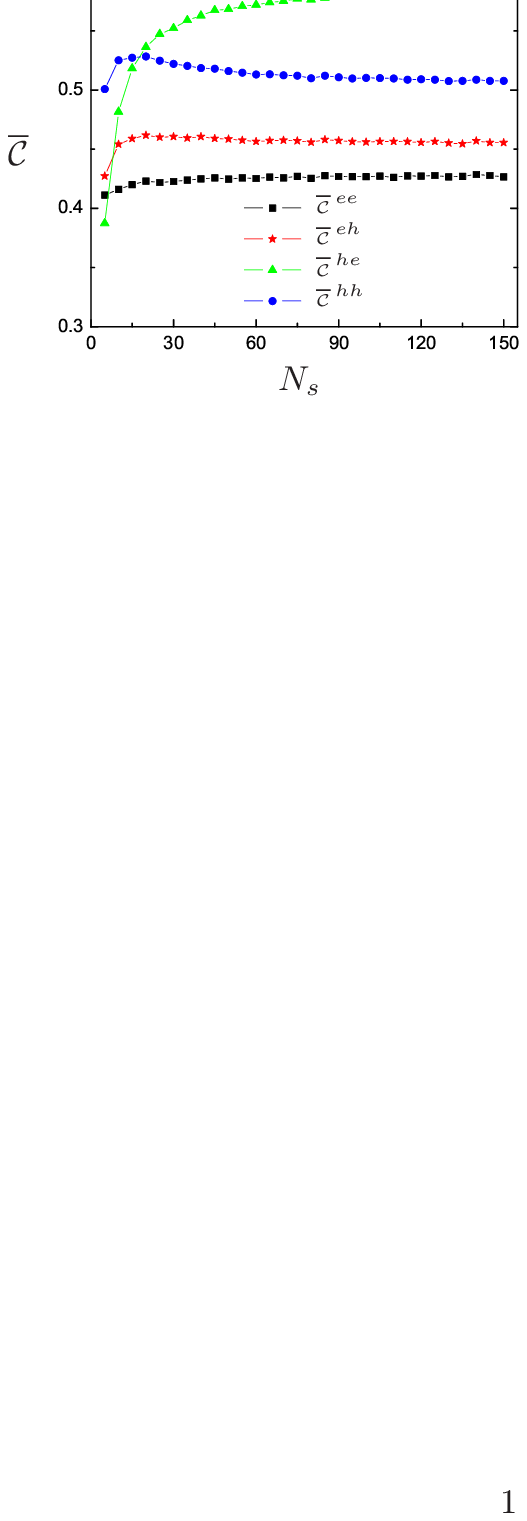}
\includegraphics[scale=1,clip]{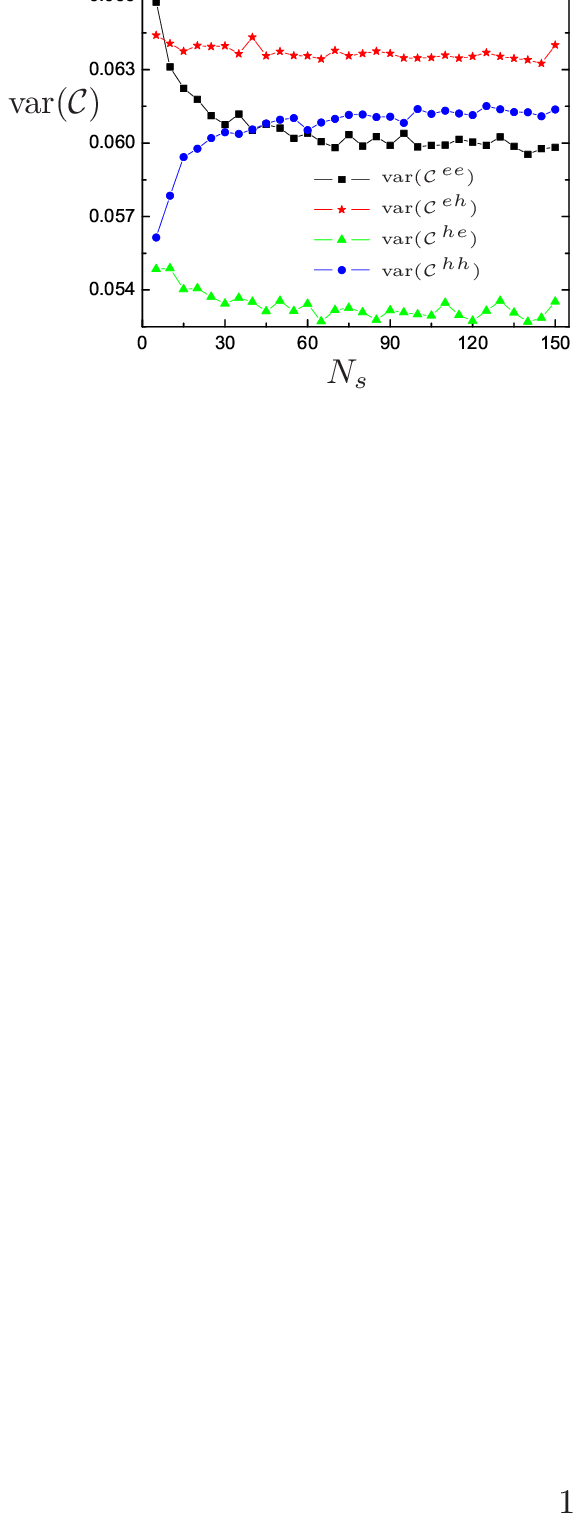}
\end{center}
\caption{(Color online) a) Mean value and b) variance of the concurrence calculated from
$10^5$ random matrices for each value of $N_s$.} \label{fig.mean.var.concurrence}
\end{figure}

The concurrence of the entangled states is given by\cite{wootters,Beenakker.Hall} \be
\mathcal{C}^{\alpha\beta}=2\sqrt{\mathrm{Det}(\gamma^{\alpha\beta}{\gamma^{\alpha\beta}}^\dagger)}/
\mathrm{Tr}(\gamma^{\alpha\beta}{\gamma^{\alpha\beta}}^\dagger).\label{eq.concurrence}
\ee In contrast with the result obtained for a generic normal
system,\cite{beenakker.emaranhador} this expression depends not only on the eigenvalues
of the normal scattering matrix, but on its eigenvectors as well. This prevents us from
proving the geometric constraint for the squared norm and the concurrence discovered in Ref. \onlinecite{Sergio-Marcel-normal} for normal systems, that forces
$\N(1+\C)< 1$. Nevertheless, we have observed in numerical simulations that this
constraint seems to remain valid for the considered system. Using Eqs.
(\ref{eq.gamma.matrix}), (\ref{eq.concurrence}), and taking into account the unitarity of $S_{NS}$, a weaker constraint can be proved,  \be \N^{\alpha\beta} \C^{\alpha\beta} \leq 1/2.
\label{eq.constriction.new} \ee This inequality is not related to specific physical
properties of the system, but it is an implication of unitarity and the design of the
device.

Fig. \ref{fig.dist.concurrence} shows the distributions of concurrence for $N_s=5$, $25$,
and $100$. They suggest that all pairs of left-right outgoing quasi-particles are
entangled, but none of them is maximally entangled. It is clearly seen that $\C^{he}$ is
the quantity most sensitive to the presence of the superconductor. However, variations
are small for all the distributions when $N_s$ is large, which indicates convergence, as
can be observed in Fig. \ref{fig.mean.var.concurrence} where we show the average value
and the variance of the concurrence.

As the number $N_s$ increases, there is a remarkable enhancement in the production of
entanglement for the hole-electron component: $\overline{\C}^{he}$ reaches values above
$0.58$, reflecting a considerable increase compared to the normal case, where
$\overline{\C} \approx 0.38$.\cite{beenakker.emaranhador} Note that the other three
components have only slight dependence on $N_s$. Moreover, variations of $\C^{he}$ from sample to sample are more concentrated around its mean value than for the other components. The variance of this quantity decreases as $N_s$ increases (see Fig. \ref{fig.mean.var.concurrence}b). We found that $\mathrm{var}(\C^{he})$ oscillates around $0.53$ for large $N_s$. This value is slightly less than the variance for the other components, as well as those found for a normal cavity.\cite{beenakker.emaranhador}

\section{Summary and conclusions}\label{sec.conclusions}

We studied orbital entanglement production in Andreev billiards, assuming time-reversal
symmetry in the system. Numerical simulations were performed to statistically analyze the
squared norm and the concurrence of states describing a pair of outgoing quasi-particles.
The behavior of the squared norm for varying numbers of channels in the superconducting
lead was explained as a consequence of Andreev reflection. Even though the norms of the
entangled components decrease as the influence of the superconductor is stronger, the
squared norm of the state describing one left-outgoing hole and one right-outgoing
electron always dominates. In this regime, there is a notable enhance in the amount of
entanglement for that state. The mean concurrence reaches values greater than $0.58$,
exceeding the value found for a normal cavity.

There exist a general feature in the entanglement production using the geometry studied
in this paper and other works.\cite{beenakker.emaranhador,Frustalgia,gopar,chico,vivo,Sergio-Marcel-normal} The
profit obtained in the amount of entanglement is, in some way, compensated by a decreased
probability of producing the state. In the case of the system under consideration, this
tendency is reinforced by the proximity effect.

It is still an open problem to establish a suitable setup for detection of the
entanglement carried by each type of outgoing pair of quasi-particles.

\begin{acknowledgments} This work was supported by FAPESP.
\end{acknowledgments}

\appendix

\section{Equations to perform the numerical simulations}\label{sec.simulations}

Let's define the matrix of the normal cavity as \be
    S_{0}=\left(
    \begin{array}{cc}
        r_0 & t^{\prime}_0\\
        t_0 & r^{\prime}_0
    \end{array}
    \right).
    \label{eq.S0.matrix}
\ee Our numerical simulations are performed through random generations of $S_0$, which
has dimensions $2 N_s \times 2 N_s$. We need to close $N_s-4$ conduction channels in the
normal side (assuming $N_s > 4$), which is done introducing a fictitious barrier
with very small transparency. Only four channels remain open, one for each normal lead.
The scattering matrix for the fictitious connector is given by \be
    S_{c}=\left(
    \begin{array}{cc}
        r_c & t^{\prime}_c\\
        t_c & r^{\prime}_c
    \end{array}
    \right).
    \label{eq.Sc.matrix}
\ee We choose $r_c=r^{\prime}_c$, $t_c=t^{\prime}_c$, and define \be
    r_{c}=\left(
    \begin{array}{cc}
        \sqrt{\Gamma}I_4 & 0\\
        0 & \sqrt{1-\Gamma}I_{N_s-4}
    \end{array}
    \right),
    \label{eq.rc.matrix}
\ee and \be
    t_{c}=\mathrm{i}\left(
    \begin{array}{cc}
        \sqrt{1-\Gamma}I_4 & 0\\
        0 & \sqrt{\Gamma}I_{N_s-4}
    \end{array}
    \right),
    \label{eq.rc.matrix}
\ee $\Gamma$ being the transparency of the barrier, and $I_n$ the $n \times n$ identity
matrix.

First, it is necessary to compose $S_0$ and $S_c$, to obtain the total normal scattering
matrix $S_N$. This is done using the following equations:
\begin{align}\label{eq.S_N.blocks}
r_N  &=  r_c + t_c M_N r_0 t_c,  \\
t_N  &=  \mathrm{i} \; t_0 (I_n + r_c M_N r_0) t_c,\\
r^\prime_N  &=  -r^\prime_0 - t_0 r_c M_N t^\prime_0, \\
t^\prime_N  &=  \mathrm{i} \; t_c M_N t^\prime_0,
\end{align}
where $M_N = (I_{N_s} - r_0 r_c)^{-1}$. The phase of the superconductor is irrelevant and
it can be set equal to zero. Neglecting the energy-dependence, the blocks of $S_{NS}$ in
the electron-hole space can be written as \cite{rev.beenakker}
\begin{align}
S_{NS}^{ee} =  r_N - t^\prime_N (r^\prime_N)^{\ast} M_e t_N,\label{eq.S_NS^ee.matrix}\\
S_{NS}^{eh} =  - \mathrm{i}\; (t^{\prime}_N)^{\ast} M_e t_N,\label{eq.S_NS^eh.matrix}
\end{align}
where $M_e=(I_{N_s} + r^\prime_N (r^\prime_N)^{\ast})^{-1}$. On the other hand,
$S_{NS}^{\;\alpha\beta}$ is structured as \be
    S_{NS}^{\;\alpha\beta}=\left(
    \begin{array}{cc}
        r^{\alpha\beta} & {t^{\prime}}^{\alpha\beta}\\
        t^{\alpha\beta} & {r^{\prime}}^{\alpha\beta}
    \end{array}
    \right).
    \label{eq.S_NS^alpha,beta.matrix}
\ee $S_{NS}^{\;eh}$ and $S_{NS}^{\;hh}$ are not necessary to calculate norm and
concurrence, because we consider the initial state given by two electrons entering the
cavity from the left.

\section{Some analytical considerations about $\N^{\alpha\beta}$}\label{sec.diagrammatics}

\begin{figure}
\begin{center}
\includegraphics[scale=0.5,clip]{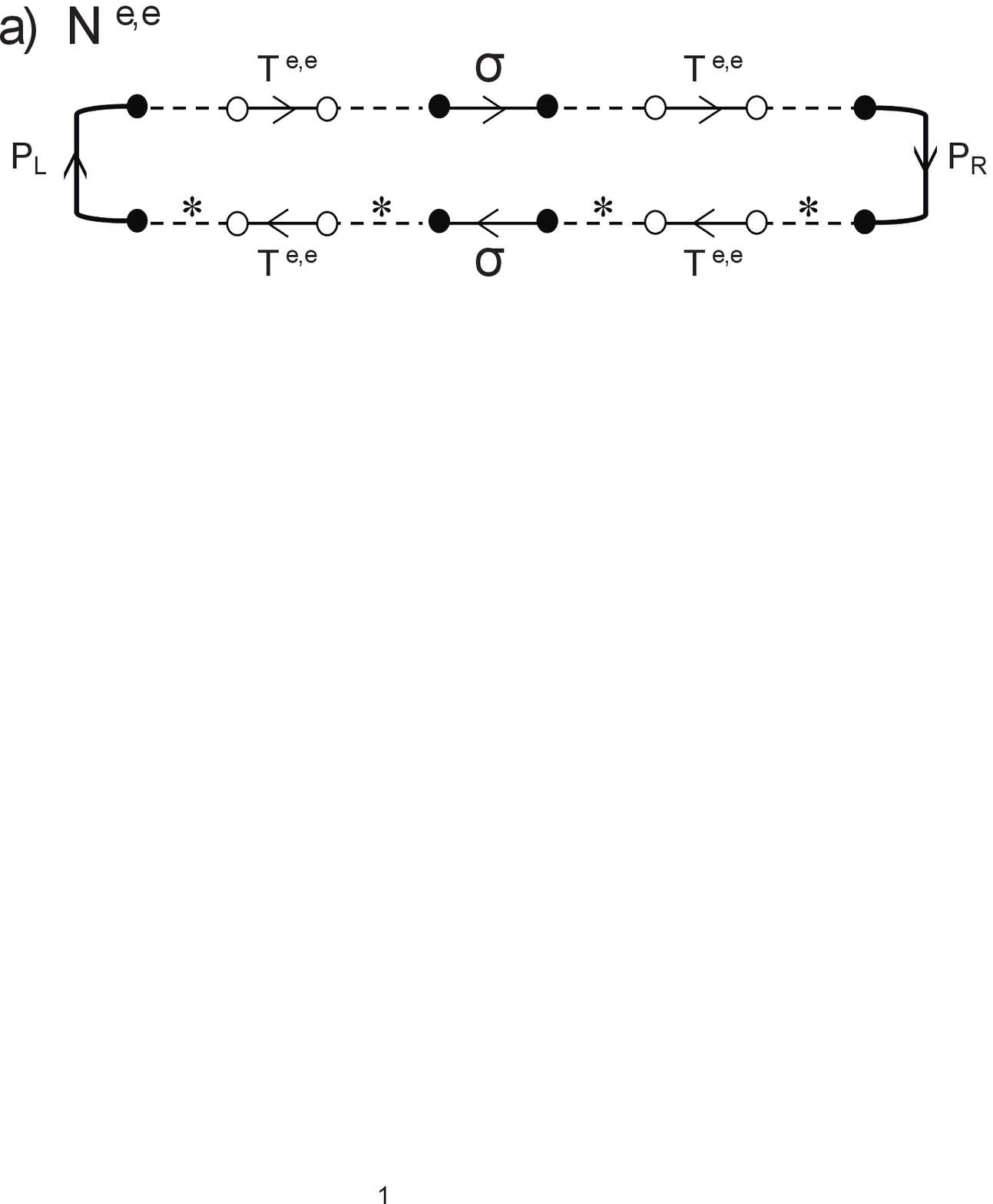}
\includegraphics[scale=0.5,clip]{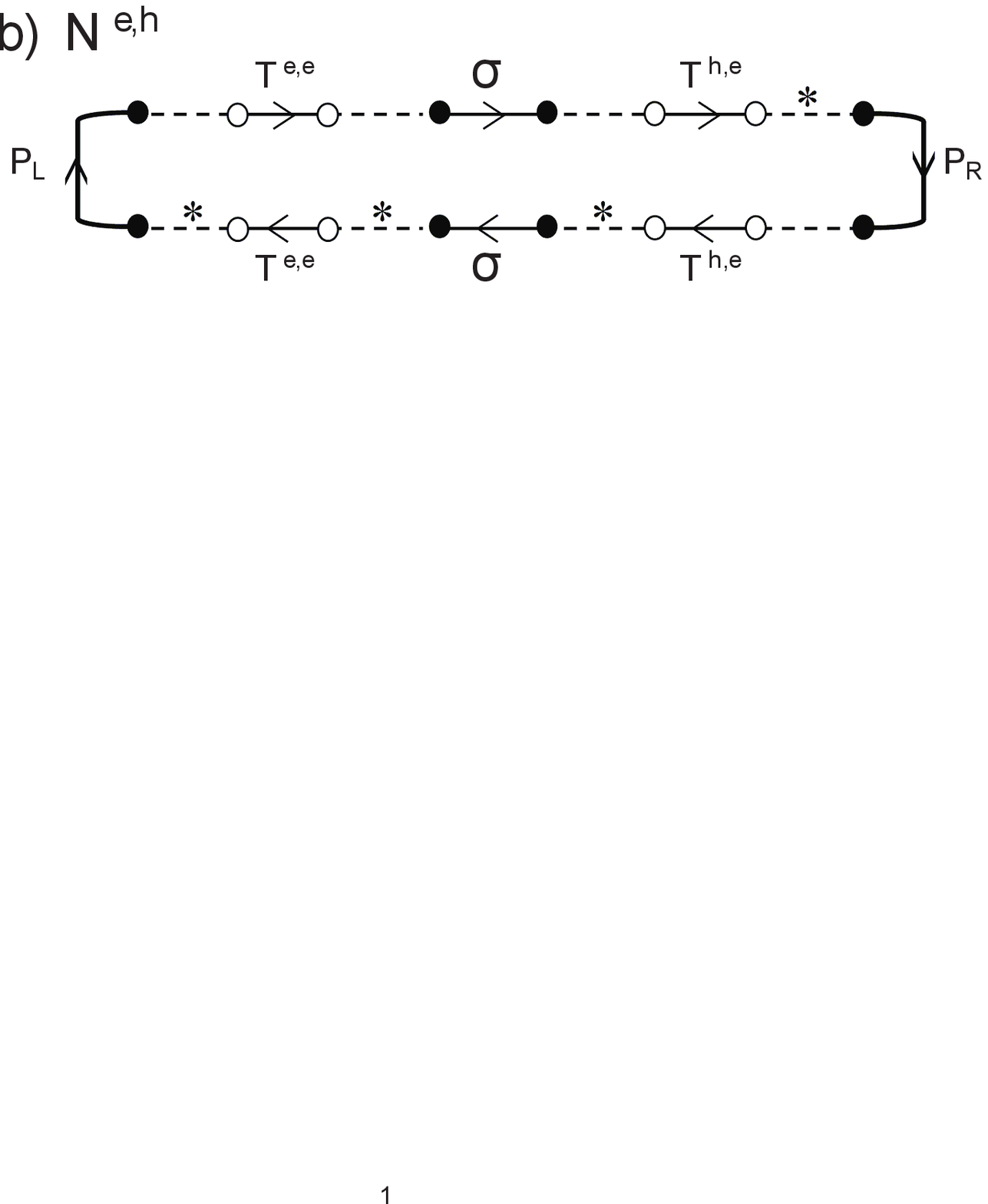}
\includegraphics[scale=0.5,clip]{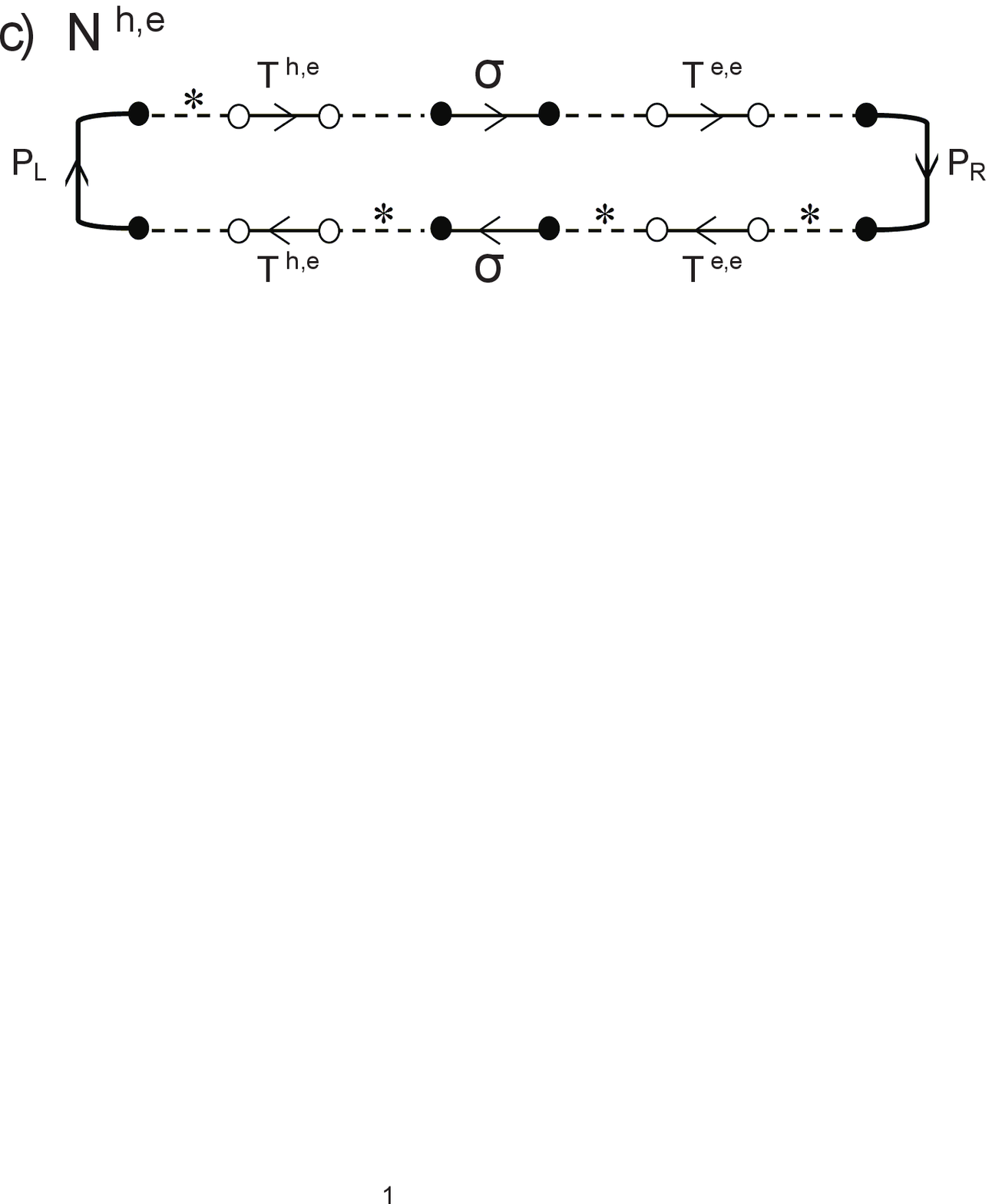}
\includegraphics[scale=0.5,clip]{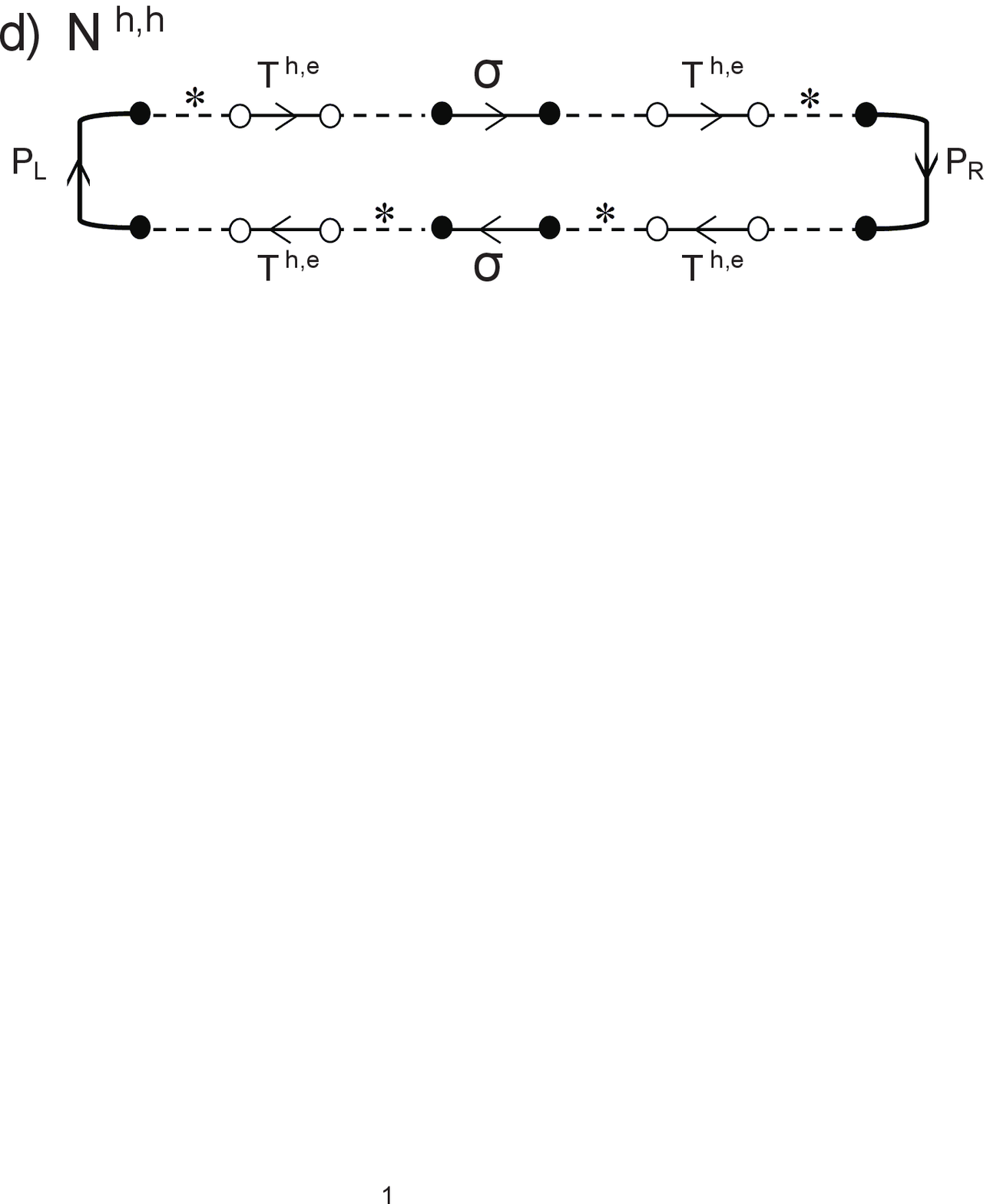}
\end{center}
\caption{The squared norms of the entangled components are sketched using the symbology
of the diagrammatical method for averaging over the unitary group.} \label{fig.diagrams}
\end{figure}

We proceed analogously to Ref. \onlinecite{Samuelsson(2002)}, where a similar device was
studied. However, an essential difference is the small numbers of open channels in the
normal leads, which requires to take into account all diagrams.

It is convenient to parameterize the scattering matrix of the cavity through the polar
decomposition \cite{mello-pereira-kumar(1988),rev.beenakker} \be S_0=\left(
    \begin{array}{cc}
        v \sqrt{I_4 - T} v^T & v \begin{pmatrix} 0 & \mathrm{i} \sqrt{T} \end{pmatrix} w^T\\ \\
        w \begin{pmatrix}0 \\ \mathrm{i} \sqrt{T}
\end{pmatrix} v^T \;\;& \;\;w \begin{pmatrix} I_{N_s-4} & 0\\
0 & \sqrt{I_4 - T} \end{pmatrix} w^T
\end{array} \right). \label{eq.polar.decomposition}
\ee In the above equation $v$ and $w$ have dimensions $4 \times 4$ and $N_s \times N_s$,
respectively, and belong to unitary circular ensembles. They account for the isotropy
assumption.\cite{rev.beenakker} On the other hand, $T={\rm diag}(\tau_{L_1},
\tau_{L_2},\tau_{R_1},\tau_{R_2})$ is a function of the transmission eigenvalues. The
eigenvalue $\tau_{L_j(R_j)}$ quantifies the probability that an electron entering from lead
$L_j$ ($R_j$) meets the superconductor before leaving the cavity.

Making $S_N=S_0$ in Eqs. (\ref{eq.S_NS^ee.matrix}) and (\ref{eq.S_NS^eh.matrix}), and using the polar decomposition, the equations for the squared norms of the entangled components can be
written as \be \mathcal{N}^{ee} = -\mathrm{Tr} \left [ P_L v \mathcal{T}^{ee} v^T \sigma
v \mathcal{T}^{ee} v^T P_R v^\ast \mathcal{T}^{ee} v^\dagger \sigma v^\ast
\mathcal{T}^{ee} v^\dagger \right ],\label{eq.norma.ee.explicita} \ee

\be \mathcal{N}^{eh} = -\mathrm{Tr} \left [ P_L v \mathcal{T}^{ee} v^T \sigma v
\mathcal{T}^{he} v^\dagger P_R v \mathcal{T}^{he} v^\dagger \sigma v^\ast
\mathcal{T}^{ee} v^\dagger \right ],\label{eq.norma.eh.explicita} \ee

\be \mathcal{N}^{he} = -\mathrm{Tr} \left [ P_L v^\ast \mathcal{T}^{he} v^T \sigma v
\mathcal{T}^{ee} v^T P_R v^\ast \mathcal{T}^{ee} v^\dagger \sigma v^\ast \mathcal{T}^{he}
v^T \right ],\label{eq.norma.he.explicita} \ee

\be \mathcal{N}^{hh} = -\mathrm{Tr} \left [ P_L  v^\ast \mathcal{T}^{he} v^T \sigma v
\mathcal{T}^{he} v^\dagger P_R v \mathcal{T}^{he} v^\dagger \sigma v^\ast
\mathcal{T}^{he} v^T \right ],\label{eq.norma.hh.explicita} \ee where we have defined the
projectors \be
\begin{matrix}
\sigma= \mathrm{i} \begin{pmatrix}
1 & 0\\
0 & 0
\end{pmatrix} \otimes \sigma_y,\\ \\
P_L= \begin{pmatrix}
1 & 0\\
0 & 0
\end{pmatrix} \otimes I_2, \\ \\
P_R= \begin{pmatrix}
0 & 0\\
0 & 1
\end{pmatrix} \otimes I_2. \\
\end{matrix}\label{eq.projectors.matrices}
\ee $\mathcal{T}^{ee}$ and $\mathcal{T}^{he}$, on the other hand, encode the transmission
properties of the NS system. They are given by
\begin{align}\label{eq.T_alpha,e}
\mathcal{T}^{ee} & =  (2 \sqrt{I_4 - T}) (2 I_4 - T)^{-1},
\\
\mathcal{T}^{he} & = -T (2 I_4 - T)^{-1}.
\end{align}

Expressions (\ref{eq.norma.ee.explicita})-(\ref{eq.norma.hh.explicita}) can be
represented, using the symbology of the diagrammatical method for averaging over the
unitary group,\cite{math.brouwer} as shown in Fig. \ref{fig.diagrams}. Each part of the
representation indicates some kind of event. Matrices $\mathcal{T}^{\alpha e}$ connected
to two dashed lines refer to a scattering process in the NS system, meaning that an
electron enters and $\alpha$ outs. On the other hand, $P_L$ ($P_R$) selects processes
describing a quasi-particle going out at the left (right), and $\sigma$ assures that two
electrons enter from the left.

In order to make the averaging, all topologies resulting from matching white/black dots
should be generated. There are $24 \times 24=576$ different topologies for each quantity.
Every diagram contains $7$ matrices in its top, indicating a process in which two
electrons enter the cavity from the left (matrix $\sigma$), one is scattered to the left
(matrices are connected to $P_L$) and the other one to the right (matrices are connected
to $P_R$). The bottom represents the same process, but inverted in time. Note that the
chronological order of the events is determined by the sequence of matrices, and not by
the arrows.

T-cycles generated from connecting black dots indicate that most of the diagrams are
zero. If the black dots annexed to $P_R$ are matched to any other black dot, a null
diagram is generated. This happens because two events taking place at different sides of
the device can not be directly connected. For this reason, the number of diagrams
different from zero is $144$, which is still large.

Even though it is hard to calculate all the diagrams, it is possible to know their
dependence on the transmission eigenvalues, which results from T-cycles generated from
white dots. There is indeed a small number of possible functional dependences for each
squared norm, as shown bellow

\be
\N^{ee} \rightarrow \frac{(1-\tau_i)(1-\tau_j)}{(2-\tau_i)^2(2-\tau_j)^2},
\label{eq.eigenvalues.dependence.ee}
\ee
\be
\N^{eh} \rightarrow \frac{(1-\tau_i)\tau_j \tau_k}{(2-\tau_i)^2(2-\tau_j)(2-\tau_k)},
\label{eq.eigenvalues.dependence.eh}
\ee
\be
\N^{hh} \rightarrow \frac{\tau_i\tau_j \tau_k \tau_l}{(2-\tau_i)^2(2-\tau_j)(2-\tau_k)(2-\tau_l)}.
\label{eq.eigenvalues.dependence.hh}
\ee
We have assumed that $1 \le i,j,k,l \le 4$. The dependence for $\N^{he}$ is the same as the one for $\N^{eh}$.

The statistical distribution of the transmission eigenvalues is given by
\cite{Baranger.Mello,Jalabert.Pichard.Beenakker} \be P(\vec{\tau}) = C \prod\limits_{1
\leq i < j \leq 4}{|\tau_i - \tau_j|} (\tau_1 \tau_2 \tau_3 \tau_4)^{(N_s - 5)/2},
\label{eq.distribution.eigenvalues} \ee where $C$ is a normalization constant. For
increasing $N_s$, this distribution is negligible except around the point $\tau_i = 1$. Considering the scaling $\tau_i = 1 - \epsilon_i/N_s$, it is possible to find that for large $N_s$ the mean value $\overline{\N}^{ee} \sim 1/N_s^2$. The other components are proportional to $1/N_s$. In this analysis, we have neglected terms of higher order in $N_s^{-1}$, and assumed that $\overline{\N}^{hh}$ is zero in the limit $N_s \rightarrow \infty$.

It is also possible to express the squared norms as a function of the charge cumulants of a
normal device. As an example, let's do this for $\N^{ee}$. Making $\tau_i + 1/\tau_i
\approx 2$ and using the Levitov-Lesovic formula,\cite{levitov-lesovic} we find that \be
[\mathrm{Tr}(\mathcal{T}^{ee})^2]^2 \approx 4 (q_2 - q_3)^2, \ee and \be
\mathrm{Tr}({\mathcal{T}^{ee}}^4) \approx 2 [6 q_2 - 5 (q_3 + q_4 - q_5) - q_6]/15. \ee
We have denoted with $q_j$ the cumulant of transmitted charge of order
$j$, for a normal device consisting of a chaotic cavity connected to two leads with $4$
and $N_s$ open channels. Following the same argument, concurrence might also be written
as a function of the charge cumulants.

\end{document}